\documentstyle[12pt,psfig]{article}
\begin{document}
\setcounter{page}{1}

  \hspace{16.6em}
 {\Large   SUNY -  NTG    97-1
  }
\vspace*{0.8cm}

  \begin{center}
{\bf  \large                           
 Soft transverse expansion  in Pb(158~AGeV) on Pb collisions:
             \\
  preequilibrium motion or 1st order phase transition?
             \\
       }
\vspace*{0.4cm}\ \\
           H.\ Sorge
 \footnote{
  E-mail: sorge@nuclear.physics.sunysb.edu
          }
\vspace*{0.4cm}\ \\
 Physics  Department,
\vspace*{0.4cm}\ \\
  State University of New York at Stony Brook, NY 11794-3800

  \end{center}

\begin{abstract}               
Transverse expansion of centrally produced matter in Pb on Pb  collisions at 
beam energies around 158 AGeV appears to be rather `soft'. Two possible reasons 
-- an extended preequilibrium stage and a first order phase transition from a 
quark-gluon-plasma into hadronic matter --   are discussed. The softening of 
transverse expansion caused by preequilibrium dynamics is estimated with the aid
of the transport model RQMD which does not contain a first order phase 
transition. It is found that the anisotropy of transverse flow in non-central 
reactions is very different in the preequilibrium and hydrodynamic scenarios 
even if the latter are based on a strong 1st order transition.
\end{abstract}  

\newpage 
The analysis of  collective flows in ultrarelativistic
 nucleus-nucleus collisions 
 is one of the major tools  to      
study the Equation of State (EoS) of strongly interacting matter,
in particular the phase transition from a
 quark-gluon plasma (QGP) into hadronic matter
 \cite{transhydro}.
 In this Letter,
 I address the question how to distinguish 
 mechanisms which  soften the pressure 
 and therefore the  expansion in 
 the central rapidity region of ultrarelativistic
 nucleus-nucleus collisions. 
  I  am going to focus here on transverse expansion,    
  because  longitudinal flows  are presumably 
  dominated by the primordial motion of the in-going projectile and target
  nuclei. I will discuss 
 effects  due to 
 an extended preequilibrium stage
  as opposed to 
 a first order phase transition from a thermal QGP into a hadron gas. 
 Addressing possible causes of  `softness' 
 implies that there are some hints of its presence
 in ultrarelativistic nucleus-nucleus collisions.
 Indeed, analysis of experimental data reveals that the
   transverse momenta of hadrons level off 
  in heavy ion collisions between beam energies of 10~AGeV
   (at AGS)  and 200~AGeV (at CERN) \cite{nuxu96}.
 The observation  means in turn that the underlying 
 collective motion does not
 increase sizably.
 On the other side, the total energy which is dumped into the
 central rapidity region is approximately  70 percent larger at
 the higher beam energy.
 
 A first-order phase transition is generically associated
 with the presence of a softest point in the EoS.
 The tendency of matter to expand on account of its internal pressure is
 reduced  in the transition region \cite{shuryak79,hung95,rigyu96}.
 The formation of a mixed phase 
  seems therefore a natural candidate  for an explanation
 if  matter  expands  softly. 
  The preequilibrium stage of nucleus-nucleus collisions is characterized
 by strong damping of transverse motion as well.
 How can the two sources of softness be distinguished?
  Here I suggest to utilize the different time 
  orderings of when the expansion of matter  becomes soft in the two cases.
     Preequilibrium softening is bound to
  happen initially. In contrast, the softness of the mixed phase evolution may
  be preceded by a fast evolution in the quark-gluon stage. A very hot QGP
  expands   presumably according to  a hard equation of state  close to
   $p=e/3$ which is valid for an ideal gas of massless particles.
   
  It is important for the idea explored here  that there are two
   transverse flow  components in 
   nucleus-nucleus collisions with non-zero impact parameter.
   At sufficiently high beam energies the  component in the reaction plane
   is expected to be stronger than the out-of-plane component.
  Anisotropies as a signature
   for hydrodynamical motion in ultrarelativistic nucleus-nucleus reactions 
 have been discussed  some time ago by Ollitrault \cite{ollitrault}.
   He suggested that a flow anisotropy is created by the anisotropy of the 
   almond-shaped initial overlap region between projectile and target.
   Hydrodynamic flows are driven  by pressure gradients  which
   are  mainly directed along the impact parameter in this case.
   One should note that this anisotropy of transverse
   flow has nothing to do  with the so-called directed flow which has been
   studied very much at lower beam energies.
  Azimuthal average  and elliptic  deformation of the two flow components
   each exhibit different degree of
   sensitivity to the early pressure.  
  It has been realized only recently that 
   the anisotropy of the flow tensor is especially  sensitive
   to the {\em early }  pressure  \cite{sorprl97}.
   It is suggested here that 
   measuring  the two flow components separately 
   will provide vital  information about the timing of the
  softening in $AA$ collisions at energies around and even larger than 160AGeV,
   e.g.\ at RHIC.
  Preequilibrium dynamics does not spoil the usefulness
  of such kind of analysis.  Quite to the contrary, analysis of 
  azimuthal asymmetries may be used to gain insight about the duration
  of preequilibrium motion.
   
 The appropriate tool to address  preequilibrium phenomena 
 in nucleus-nucleus collisions is transport theory. Here I am employing the
 relativistic quantum molecular dynamics (RQMD) approach \cite{SOR95}.
  In this Letter, I am going to study  one particular reaction,
   Pb(159AGeV) on Pb collisions. Such reactions  
    are currently explored by various
   experimental groups  at CERN. 
  It should  be mentioned that the `inertial confinement' effect
  from spectators discussed in Ref.~\cite{sorprl97} 
   is very weak for the collisions at
  CERN energy, at least in the model. Sizable pressure builds up
  only after the spectators have disappeared from the central region.
  Furthermore, it can be  expected that
  heavy-ion reactions at this beam energy create energy densities in
  the central region which may be sufficient for QGP formation.
   This Letter  should be viewed as complementary to
  the analysis presented  in Ref.~\cite{sorplb96}. Corrections to the
    ideal hydrodynamic evolution  in the 
 later dilute stages (post-equilibrium) were discussed in the earlier work.
    The motivation for the present study has been to get a  handle on
    the role of the preequilibrium stage      
   for  transverse expansion dynamics. Subtracting the nonequilibrium
   effects would enable one  to extract the `thermal' properties of the
    quark-hadron transition from experimental data.
   
 The outline of this Letter is as follows.  Using RQMD the softening 
    of  transverse expansion caused by preequilibrium effects
  is estimated. By comparing final hadron spectra 
   to recent preliminary NA49 data
  it is checked that the expansion from  RQMD
   for Pb on Pb collisions at CERN energy
   is compatible with experimental observations.
   The RQMD calculation provides an example of
    an expansion dynamics {\em without}
   a first order phase transition. 
   I will demonstrate that elliptic flow distinguishes
   the RQMD-type preequilibrium scenario from hydrodynamical evolution
   based on a strong first order phase transition.
    
 A detailed overview of the  RQMD model
 can be found elsewhere \cite{SOR95}. Here I summarize only       
 how the ingredients affect the transverse expansion.
 RQMD is based on string and resonance excitations in
 the primary collisions of nucleons from target and projectile.
 Overlapping color strings  fuse into ropes, flux-tubes with
  sources of larger than fundamental  color charges.
 Color strings and ropes model the prehadronic stage in 1+1 dimensions.  By 
 construction, they do not exert any 
 transverse pressure on their environment.
 Indeed, Trottier and Woloshyn have shown that according to their lattice gauge 
  simulations color ropes do not expand in transverse dimensions,
 contrary to naive bag model expectations \cite{TRO93}.
  Transverse expansion in the central region   starts only after
  hadronization, because the initially generated transverse momenta from string
  and rope fragmentation are oriented randomly.
  The hadronic
  expansion stage in RQMD starts off far from kinetic equilibrium.
  One  reason is that
   the nuclear thickness sets a minimum time  for nonequilibrium
   due to the finite crossing time of projectile and target. 
\setcounter{footnote}{0}     
\footnote{
  Sometimes the  crossing time is  unjustifiably 
  ignored  in the literature. 
   E.g.\ in Ref.~\cite{cleymans96} it is assumed that all entropy is  produced
  already after 1 fm/c  --  before the two Pb nuclei have
   even completely passed through each other. 
   The large densities which result from this choice of initial
   conditions form the base of
    the claim in Ref.~\cite{cleymans96} that a QGP is a more `natural'
    thermal state than a resonance gas at CERN energies. 
 }
    The  crossing time of the two Pb nuclei
    in an observer frame  with CMS rapidity is 1.4 fm/c at 160~AGeV 
      and  therefore  even larger than the
   hadronization time from string and rope fragmentation.
    Hadrons in the same space-time area may have very different rapidity
    initially -- just 
       because they are produced in 
    elementary collisions with different  locations along the beam axis.
  The spreading  of longitudinal velocities  $ \delta \beta$
    from the dispersion 
    of  collision points  can be easily estimated 
     in the Bjorken scenario  \cite{Bjorken} 
   (with formation time  $\tau_0$ taken  to be 1fm/c)
   which bears some
    resemblance to string-type approaches. 
   $\delta \beta$ may take values up to
  \begin{displaymath}
     \delta \beta =  
      \left(1+ \left(
        \frac{\tau _0 \gamma}{2 R_A} \right)^2 \right)^{-1/2}
       \qquad .
  \end{displaymath}
  The corresponding difference in rapidities 
   $\delta y$=1/2 $\ln ((1+\delta \beta)/(1-\delta \beta))$
    amounts to approximately 1.1 units    for Pb(159AGeV)+Pb collisions.
   This dispersion  comes on top
   of  rapidity fluctuations from hadronization
    which have the same order of magnitude.         
   The total dispersion of local hadron rapidities is clearly much larger
    than in thermal equilibrium in which the width of
    rapidity distributions is restricted not to exceed 0.7 units. 
    Initially,  
    the local momentum distributions in RQMD 
   are therefore  elongated along the beam axis.
    This  diminishes expansion in transverse direction
   in  comparison to the kinetic equilibrium case.   
   One can define an  effective  pressure $p$ employing 
     the spatial diagonal components of the energy-momentum tensor
   (see Ref.~\cite{sorplb96}). The 
    softening of the transverse expansion
   from preequilibrium anisotropies shows up as a 
   reduction of the transverse pressure.

 In the hadronic stage of RQMD, the fragmentation products from rope, string
 and resonance decays  interact with each other and the
 original nucleons, mostly via binary collisions. 
 These interactions drive the system towards equilibrium       
 \cite{sorplb96}. 
 The equilibrium state in RQMD is  an ideal gas  of
 hadrons and resonances, 
 up to small corrections from strings
 and neglecting contributions from mean-field type
  potentials between baryons (which have not been employed for the
 present study).
 The  relevant quantity for hydrodynamic expansion is the
  pressure as a function of energy density. At relevant temperatures
  around 150-180 MeV the ratio $p/e$ stays approximately around 1:6
  (for $\mu _B$=0)  if all experimentally well confirmed
 hadronic states with masses below 2 GeV/$c^2$
 are included \cite{bebie92}.  This is rather close to the
  spectrum of states included in RQMD \cite{SOR95}.
   The inclusion of resonances already strongly
   softens the EoS. A pion gas at same energy density
   would provide twice as much  pressure as the resonance gas.
  
\begin{figure}[h]

\centerline{\hbox{
\psfig{figure=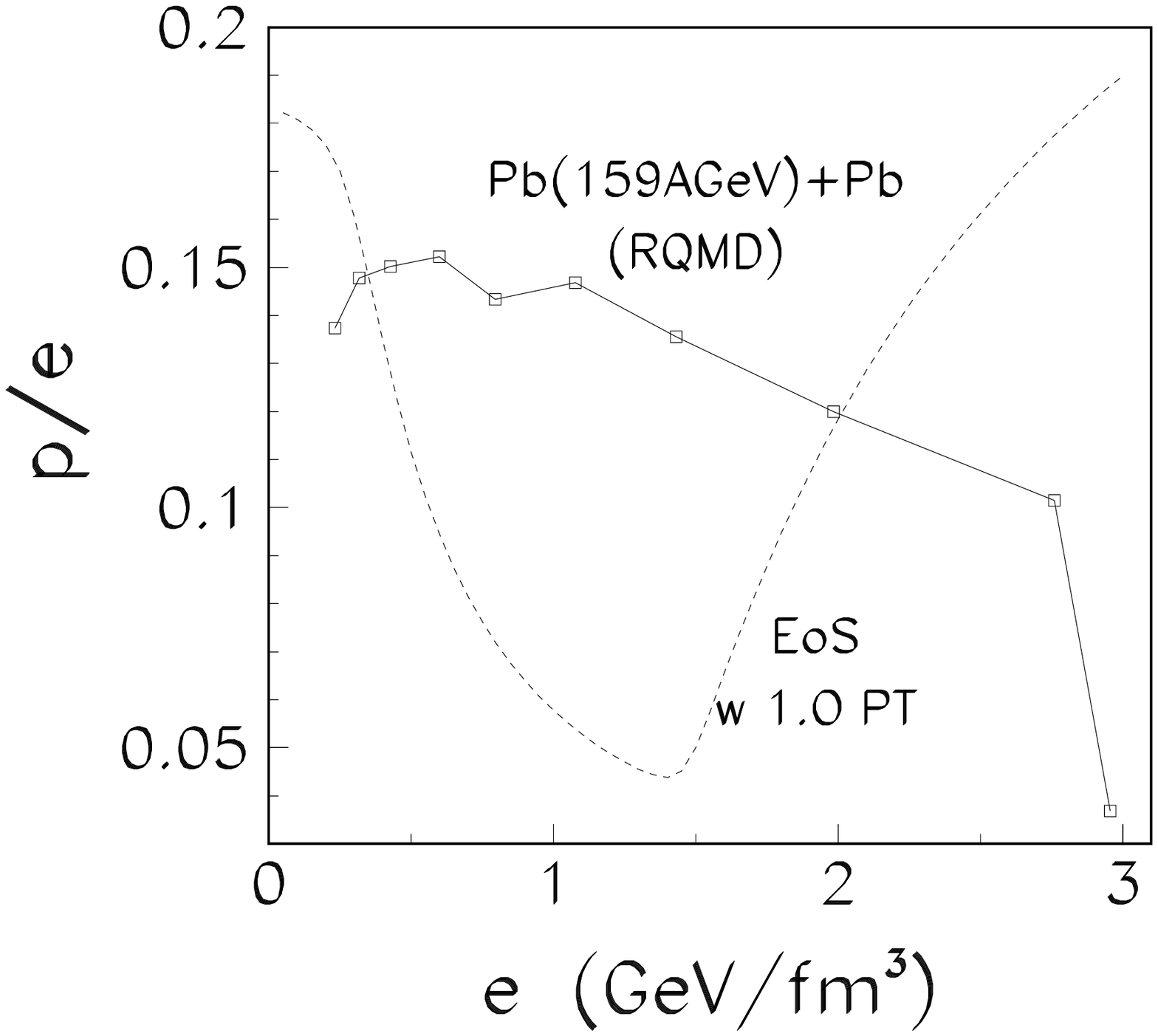,width=15cm,height=15cm}}}

\caption
[
 ]
{
 \label{epratio_rqmd-th}
}
\end{figure}

   Approximately 300 Pb(160AGeV) on Pb collisions       
    with an  impact parameter $b$=6 fm have been calculated with
    the RQMD model (version 2.3) for the present study.
   The  hadronic energy-momentum tensor
   has been evaluated in the   collision center.
 Fig.~\ref{epratio_rqmd-th} displays the time evolution of the 
  event-averaged local  transverse pressure and energy density.
   Initially, the energy density is very large, close to
    3 GeV/fm$^3$.  However, a rather large fraction of 
    this energy resides in the `hidden' collective motion
     along the beam axis. As a consequence, 
  the transverse pressure is considerably
 softened for a time interval of about 4 fm/c. 
  Usually, hydrodynamic simulations of ultrarelativistic
   collisions  assume that there is no relevant transverse
   expansion before local equilibrium has been achieved.
   This approximation can be justified only if  the equilibration time
    would be small compared to both the total interaction time and to
     the transverse size of the collision region 
    divided by $c$  \cite{ollitrault}.
   Taking the prehadronic and hadronic evolution together
  the  preequilibrium stage lasts for  approximately
   5 fm/c. 
   Such large values make  the preequilibrium
   evolution utterly relevant for the
    transverse expansion dynamics.  
    Note that $e$ and $p$ evolve rather 
     similarly in the other areas
   of the collision region although the maximum densities are
   somewhat lower than in the center.
 
  After equilibration, the ratio $p/e$ 
  approaches  values around 1:6 which are expected
   for the hadronic system in kinetic and chemical equilibrium.
   It provides indirect evidence that 
    chemical equilibrium must be close at this later stage. 
   Finally,  the $p/e$  ratio drops again signaling the break-down
   of near-equilibrium dynamics. 
    The pressure goes down faster than the
    energy density, because the dilute gas in the
     central region is characterized 
      by more  massive
      constituents. One reason is loss of chemical equilibrium.
  Finally,  the system in the center 
    becomes a  dilute gas  dominated by
   nucleons and not--  contrary to naive expectation -- by  pions.  
   Increasingly with time, interactions  are ceasing
  and being replaced  by free streaming. 
   The lighter mass particles
   leave the collision center faster than the other ones.
   
\begin{figure}[h]

\centerline{\hbox{
\psfig{figure=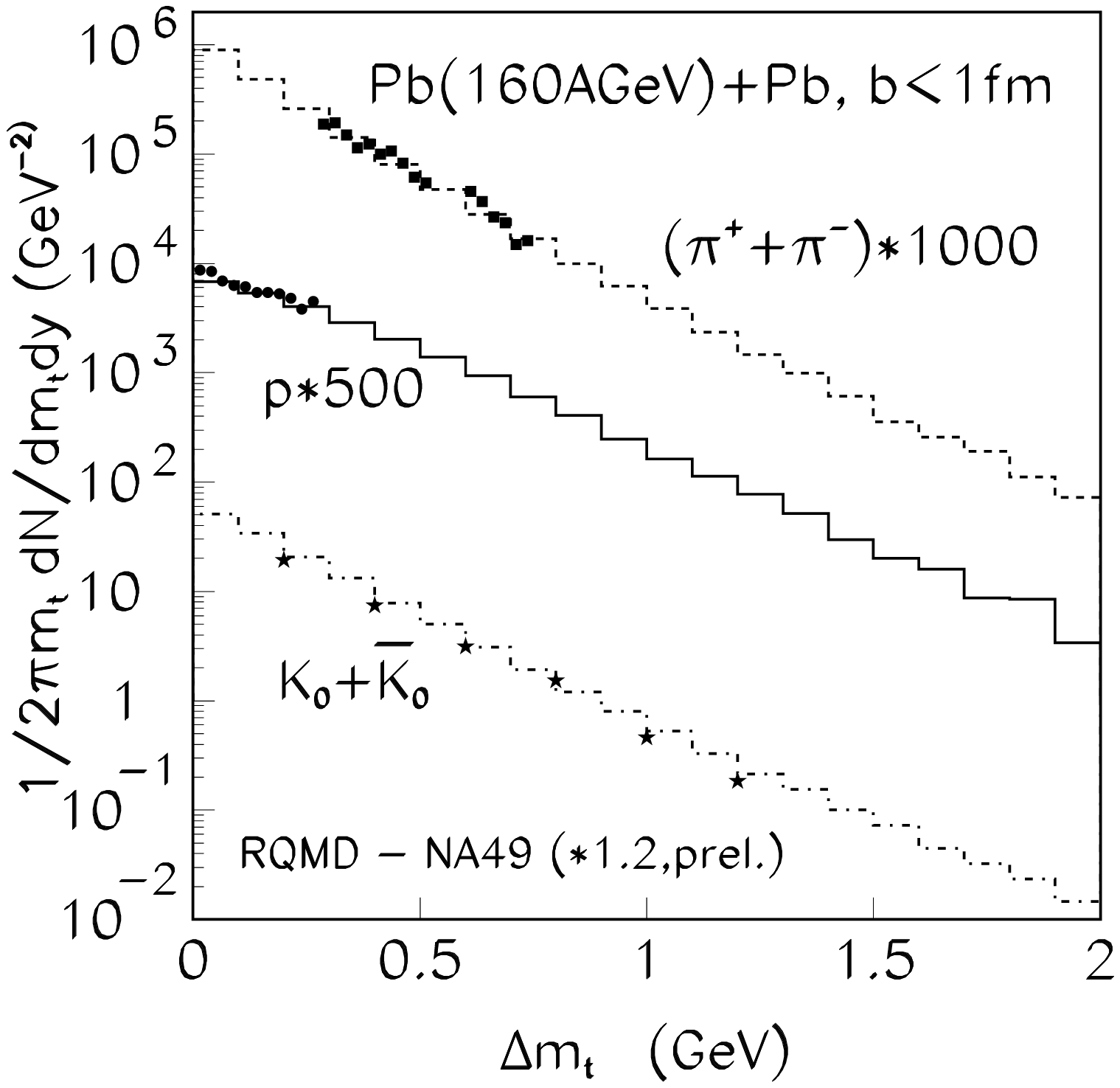,width=15cm,height=15cm}}}

\caption
[
 ]
{
 \label{pbpbtrspect}
}
\end{figure}

   Transverse momentum spectra  of
   different hadron species can be employed to
   check whether the average  transverse  (radial)
   flow generated by RQMD
   has the appropriate strength. Fortunately, the
   often discussed ambiguity in the hydrodynamic model
 concerning  trade-off between temperature and flow \cite{schneder93} which is
  related to the choice  of freeze-out criteria 
  does not exist for RQMD. Transport models like RQMD
  which have included  two-body scattering in accordance with free-space data
  become accurate in the dilute gas limit.                     
  Fig.~\ref{pbpbtrspect} contains a comparison of RQMD predictions for
  hadron spectra at central rapidity which were taken 
  from Ref.~\cite{sorplb96} with recent preliminary NA49 data \cite{na49_qm96}.
  The shapes of the calculated spectra which directly 
  reflect the flow effects 
  (and also the particle ratios)
  agree well with the data.
   
  In order to assess the importance of preequilibrium dynamics
  on the transverse expansion 
  the  `effective EoS' $p$($e$) from RQMD is compared
   to an equilibrium EoS with a 1st order phase transition (PT)
   in  Fig.~\ref{epratio_rqmd-th}.
  The equilibrium EoS
  with critical temperature at 160 MeV      
  was employed for the hydrodynamic studies
  in Ref.~\cite{hung95}.
  The  comparison  reveals
   that   preequilibrium in RQMD softens its effective EoS 
  as much as the  latent heat does for the equilibrium EoS. Thus
  it seems conceivable that the RQMD nonequilibrium dynamics without
    and a hydrodynamic evolution with a strong PT included generate 
   comparable average transverse
  flows if  initial energy densities around 2-4 GeV/fm$^3$
    are chosen.
  In fact, results of some hydrodynamic studies 
   incorporating an EoS with  phase transition
   have been published
   which  fit  measured slope parameters 
   for collisions at CERN energy  reasonably  well 
  \cite{cleymans96},\cite{dumitru95}-\cite{sollfrank96}.  
     
    How can one distinguish the equilibrium scenario with strong 
    1st order  transition from the RQMD-type preequilibrium softening
     of transverse expansion? 
    The different time ordering of hard and soft expansion stage
     in  the two cases which is visible from Fig.~\ref{epratio_rqmd-th} 
     is quite suggestive for a solution.
      One may look for observables other than radial flow which  are also
    sensitive to the  pressure, however, with different relative
    weight of early and late stage.  
    Here the anisotropies of the transverse flow in
      non-central collisions 
    are a promising candidate, because 
   they are arguably more sensitive to early pressure.
  They evolve only as far as the system retains some
 memory of the initial anisotropy, because
  the  anisotropic  overlap zone of projectile and target nucleus
  is responsible for the  asymmetries.
  Furthermore, the developing stronger in-plane flow acts against its cause,
  the asymmetry
   of the collision zone.  In contrast, 
  the late stage is weighted more heavily in the evolution of the 
   average flow. 
  The weight is essentially proportional to the system size
  (due to the $p dV$ term in the thermodynamic approximation).
   The different sensitivity of elliptic deformation and average flow
   to the earlier pressure explains 
   some  amusing results of
   hydrodynamic calculations  in Ref.~\cite{ollitrault}.
   Ollitrault has  found that
   additional  expansion in the longitudinal direction
   {\em strengthens }  the azimuthal asymmetry as compared to a 
   2-dimensional transverse expansion.
   He discusses this in terms of hydrodynamics approaching eventually
    a scaling solution faster in two than in three dimensions. 
   On the other side, the radial 
    flow is  getting weaker if another dimension for expansion
   is open. 
   Earlier freeze-out due to transport
   into longitudinal direction 
     causes these opposite trends.
    
\begin{figure}[h]

\centerline{\hbox{
\psfig{figure=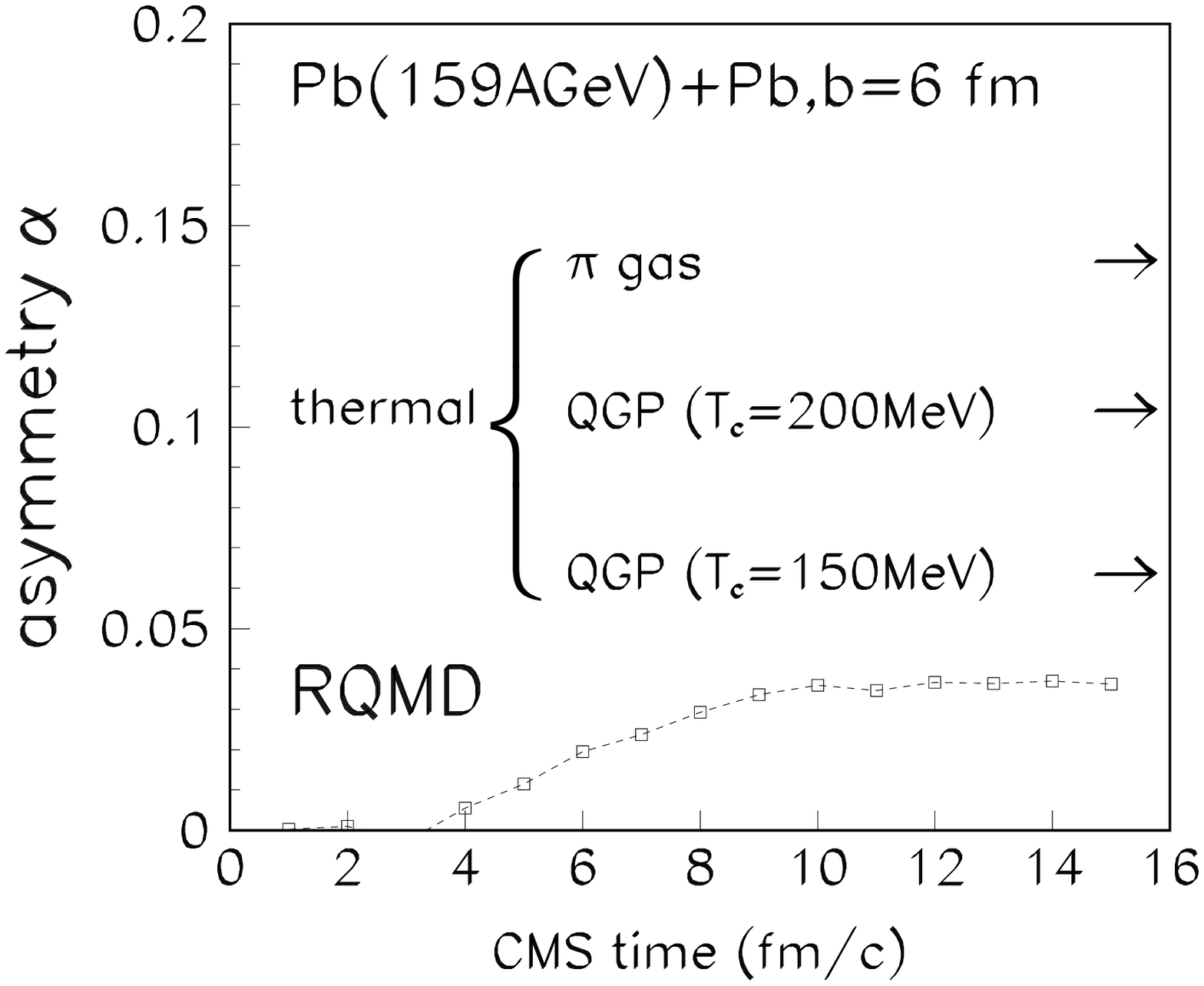,width=15cm,height=15cm}}}

\caption
[
 ]
{
 \label{asy_cern}
}
\end{figure}

    Fig.~\ref{asy_cern}  displays how the 
     azimuthal asymmetry of transverse flows develops with time
     according to the RQMD calculation.
    The azimuthal asymmetry  of transverse hadron momenta
    can be quantified by defining the
      dimensionless variable $\alpha $    via
  \begin{equation}
     \alpha =
          \frac{
             \langle  p_x^2  \rangle 
            -  \langle  p_y^2  \rangle 
               }
               {
             \langle  p_x^2  \rangle 
            +  \langle  p_y^2  \rangle 
               }
         \qquad .
   \label{eqalpha}
  \end{equation}
   $p_x$ ($p_y$) denotes the transverse momentum component of hadrons
   parallel (orthogonal) to the impact parameter vector.
    The time evolution of $\alpha $  in
     Pb(159AGeV) on Pb collisions with $b$=6~fm
    generated from RQMD 
    has been calculated and is plotted in   Fig.~\ref{asy_cern}.
     A central rapidity cut ($y_{CMS}\pm 0.7$) has been applied.
    $\alpha$($t$) stays close
  to zero in the preequilibrium stage which  reflects
  the small transverse pressure during this stage.
  Only after the system approaches kinetic equilibrium the
  flow develops a  stronger asymmetry.
 The final $\alpha$ value  is  reached after approximately 10 fm/c.
  Fig.~\ref{asy_cern} exhibits also the final values 
   of $\alpha $ for the 
 same system  but calculated with 
  boost-invariant hydrodynamics \cite{ollitrault}.
   Since  thermal pressure which drives the expansion 
   in these hydrodynamic calculations starts already at very early times 
    (1 fm/c),  the anisotropies become  much stronger  
   than in RQMD. The asymmetry parameter has been calculated by
   Ollitrault for different equations of state, of a pure $\pi$ gas
   and with 1st order phase transition into a QGP
    (and  critical temperature $T_c$ either 150 or 200 MeV).
  Comparing  the  hydrodynamic and non-equilibrium RQMD values 
    for $\alpha$ one finds that the $\pi $ gas value is larger
    by a factor 3.84 and the QGP values by factors
    1.74  (150 MeV) or 2.84 (200 MeV).
   It should be noted that the same hydrodynamic calculation
    with EoS as  in the equilibrium limit of RQMD would give an
   $\alpha $ value  between $\pi$ gas and QGP value.
  We find indeed 
   that the anisotropies are more sensitive to
    an early softening from preequilibrium  than to a later phase transition.
   
  In this paper I have discussed how to distinguish 
  whether  transverse expansion in $AA$ collisions at 160AGeV
  is softened by extended 
  preequilibrium motion or by a 1st order first transition
   from a thermal QGP into hadronic matter.
  The strength of elliptic flow 
  looks like a promising  candidate to shed
  light on this question.
 Fortunately, important progress has also been achieved recently
  on the experimental side. Several experimental collaborations
 have found signals of azimuthal asymmetries in the central collision region.
 NA49 has reported preliminary data on  elliptic transverse energy
 flow patterns for non-central  Pb(158AGeV) on Pb reactions 
  \cite{qm96-na49}. Energy flows 
 in neighboring pseudo-rapidity windows  are clearly  correlated.
  Similar correlations in photom emission (mostly $\pi ^0$ decays)
   have been observed by WA93 \cite{qm95-wa93}.
 Recent  E877 data for Au on Au collisions  
 show  that already at the lower beam energy of 11.5 AGeV the main
 flow direction is indeed parallel to the impact parameter \cite{qm96-e877}.   
 First  comparisons to models indicate  that the experimental signals 
 may be strong enough 
 to distinguish between different scenarios \cite{qm96-na49}.
 Hopefully,  in the near future
  the combined experimental and theoretical analysis of
 elliptic flow  will  considerably
  narrow down  the range of `allowed' expansion scenarios and
  implicitly of the EoS in the quark-hadron
 transition region. 
      
The author thanks M.\ Hung for providing the input data to the
equilibrium EoS which has been  displayed in Fig.~\ref{epratio_rqmd-th}.
This work has been
supported by DOE grant No. DE-FG02-88ER40388.

\newpage

{\noindent  \LARGE   Figure Captions:}
\vspace{0.6cm}

{\noindent \large Figure 1: }

{\noindent        
Time evolution of local transverse pressure $p$ and energy
density $e$ in the collision center 
 of  the system Pb(159AGeV) on Pb  with impact parameter 6 fm.
 Symmetry under reflections enforces that
 the spatial components of the (energy and baryon) flow velocities are zero
 at this point.
 $p$ is determined as the average of the spatial transverse components
  in the diagonal of the hadronic stress tensor. 
These quantities have been extracted
 from approximately 300 RQMD events.
 Square symbols represent calculated $p$/$e$ values versus $e$
   starting 1 fm/c after initial touching of the two Pb nuclei
   in time intervals of
   1 fm/c each (in the CMS of the two nuclei).
  The straight line is only to guide the eye.
  The dashed line represents the equilibrium  equation of state
   which was used  for hydrodynamic studies in Ref.~\cite{hung95}. 
   This EoS contains a  1st order phase transition
    with $T_c$=160 MeV.       
}
\vspace{0.2cm}

{\noindent \large Figure 2: }

{\noindent        
  RQMD prediction for the
  transverse mass spectrum   1/2$\pi m_t$ $d^2N/dm_tdy$
     at central rapidity 
   in  the reaction  Pb(160AGeV) on Pb 
   as a function of $\Delta$$m_t$=$m_t$-$m_0$: results have been 
    taken from Ref.~\cite{sorplb96} and are compared with preliminary 
   NA49 data \cite{na49_qm96}.
    The histograms represent the RQMD spectra 
    for protons (straight line),  neutral (anti-)kaons 
    (dashed-dotted line) and  
    charged pions (dashed line), the symbols
 the data for protons (full circles), 2$*K_{short}$ (stars)
 and 2$*\pi ^+$ (squares). In addition, the data have been multiplied with
   1.2 which probably reflects the tighter centrality cut
   (b$<$ 1 fm) used in the calculation.
}
\vspace{0.2cm}

{\noindent \large Figure 3: }

{\noindent        
  Time evolution of transverse momentum anisotropy parameter
   $\alpha $ calculated with RQMD for the same system as in
   Fig.~\ref{epratio_rqmd-th}.
  The variable $\alpha$ is defined in  eq.~(\ref{eqalpha}).
   Only hadrons with
    rapidities in the window $y_{CMS}\pm 0.7$ are included.
   The squares represent the RQMD values, the arrows
   the final $\alpha $ values from hydrodynamic calculations
    for the same system and three different EoS
   which have been published in the first of Refs.~\cite{ollitrault}.
}

\newpage

\end{document}